# Polar and magnetic order in GaV$_4$Se$_8$

E. Ruff,[1] A. Butykai,[2,3] K. Geirhos,[1] S. Widmann,[1] V. Tsurkan,[1,4] E. Stefanet,[4] I. Kézsmárki,[1,2,3] A. Loidl,[1] and P. Lunkenheimer[1,*]

[1]*Experimental Physics V, Center for Electronic Correlations and Magnetism, University of Augsburg, 86159 Augsburg, Germany*
[2]*Department of Physics, Budapest University of Technology and Economics, 1111 Budapest, Hungary*
[3]*MTA-BME Lendület Magneto-optical Spectroscopy Research Group, 1111 Budapest, Hungary*
[4]*Institute of Applied Physics, Academy of Sciences of Moldova, Chisinau MD-2028, Republic of Moldova*

In the present work, we provide results from specific heat, magnetic susceptibility, dielectric constant, ac conductivity, and electrical polarization measurements performed on the lacunar spinel GaV$_4$Se$_8$. With decreasing temperature, we observe a transition from the paraelectric and paramagnetic cubic state into a polar, probably ferroelectric state at 42 K followed by magnetic ordering at 18 K. The polar transition is likely driven by the Jahn-Teller effect due to the degeneracy of the V$_4$ cluster orbitals. The excess polarization arising in the magnetic phase indicates considerable magnetoelectric coupling. Overall, the behavior of GaV$_4$Se$_8$ in many respects is similar to that of the skyrmion host GaV$_4$S$_8$, exhibiting a complex interplay of orbital, spin, lattice, and polar degrees of freedom. However, its dielectric behavior at the polar transition markedly differs from that of the Jahn-Teller driven ferroelectric GeV$_4$S$_8$, which can be ascribed to the dissimilar electronic structure of the Ge compound.

## I. INTRODUCTION

In conventional ferroelectrics, the occurrence of polar order is caused by the displacement of ions into off-symmetry positions (displacive ferroelectricity) or the long-range ordering of permanent dipolar moments (order-disorder ferroelectricity), both arising at structural phase transitions. However, in recent years unconventional ways of generating ferroelectricity have come into the focus of interest, especially as they often favor the generation of multiferroic states where ferroelectricity coexists with ferro-, antiferro- or ferrimagnetism in a single phase [1,2,3,4,5]. Multiferroics have high technological potential in realizing novel types of magnetic memory devices [2,3,6] and optical diodes [7,8] due to the strong static and dynamic magnetoelectric effects characteristic of them [2].

A prominent example are the multiferroic perovskite manganites, where improper ferroelectricity is induced at the onset of helical spin order via the inverse Dzyaloshinskii-Moriya or spin-current mechanism [1,5,9]. They are so-called type-II multiferroics [10] where improper ferroelectricity emerges due to the magnetic ordering. In contrast, in type-I multiferroics, the magnetic order develops independent of the ferroelectric polarization [10]. The ideal multiferroic material hosts ferromagnetism and ferroelectricity, strongly coupled to each other. However, most of the multiferroics known up to date are antiferromagnets or ferrimagnets as it is the case for the archetypical type-I multiferroic BiFeO$_3$ [6,11,12,13].

There is a representative class of unconventional ferroelectrics, where the polarization is predominantly induced by the electronic charge and/or orbital degrees of freedom and atomic displacements play only a secondary role. For example, charge-order induced electronic ferroelectricity was found at the Verwey transition of multiferroic magnetite [14] and in various organic charge-transfer salts [10,15,16,17,18] also including multiferroic materials [16,17]. Moreover, orbitally-driven ferroelectricity was observed upon the Jahn-Teller (JT) transitions of several materials where the JT distortion is believed to trigger the ferroelectric polarization, partly of electronic and partly of ionic nature [19,20,21,22,23,24]. Especially, recently such JT-driven ferroelectricity was reported in the lacunar spinels GeV$_4$S$_8$ [23] and GaV$_4$S$_8$ [25,26] upon their JT transitions at $T_{JT} = 32$ K and 44 K, respectively. In GaV$_4$S$_8$ the characteristics of the ferroelectric-ferroelastic domains were studied by piezo force microscopy [26]. Both compounds are type-I multiferroics as they undergo magnetic phase transitions deep in the ferroelectric phase. In the ground state, GeV$_4$S$_8$ is an antiferromagnet [23,27,28,29,30,31] while GaV$_4$S$_8$ is a ferromagnet [32,33]. Moreover, for GaV$_4$S$_8$ a complex magnetic phase diagram was reported, including a cycloidal and a Néel-type skyrmion-lattice phase underlying the ferromagnetic state [32]. In addition to the ferroelectric polarization arising at the JT transition, spin-driven excess polarization was detected in all magnetic phases of this material. Notably, this also includes the skyrmion phase and it was proposed that the magnetic skyrmions in GaV$_4$S$_8$ are dressed with local electric polarization [25].

In the past, lacunar spinels also have attracted interest due to the presence of tetrahedral $M_4$ metal clusters in the lattice (with $M = $ V in the present case) that behave like molecular magnets characterized by cluster orbital and spin degrees of freedom [33,34,35,36,37,38]. It is the spin associated with these clusters that produces magnetic ordering in these systems at low temperatures. The JT transition in these compounds is driven by the degeneracy of the cluster orbitals

---

*[*]peter.lunkenheimer@physik.uni-augsburg.de

[24], leading to a distortion of the $M_4$ tetrahedra and finally inducing ferroelectric order.

At ferroelectric transitions, the dielectric properties usually exhibit strong anomalies, which indeed is also the case for GeV$_4$S$_8$ and GaV$_4$S$_8$ [23,25]. However, despite their structural similarity and the common origin of ferroelectric ordering, they behave quite differently: In GeV$_4$S$_8$ [23] the dielectric constant $\varepsilon'(T)$ exhibits a strong increase below $T_{JT}$ and shows a broad maximum several K below the transition, whose amplitude is significantly different upon cooling and heating. When approaching the magnetic transition, $\varepsilon'$ becomes smaller again. Moreover, in Ref. [29] an additional peak close to $T_{JT}$ was reported for this compound. This finding was suggested to indicate a decoupling of the JT and polar phase transitions into two subsequent transitions [29]. Overall, the dielectric behavior of GeV$_4$S$_8$ at its ferroelectric transition is unusual, which may be in line with the non-canonical nature of the polar ordering mechanism. However, in contrast to GeV$_4$S$_8$, for GaV$_4$S$_8$ a single peak right at $T_{JT}$ was found, quite similar to the behavior in canonical ferroelectrics [25,39]. This peak becomes strongly suppressed at high frequencies beyond 1 MHz, which is consistent with the expectation of relaxational dynamics for order-disorder ferroelectrics [40]. An order-disorder character of the ferroelectric transition in GaV$_4$S$_8$ is also confirmed by very recent Raman- and infrared-spectroscopy results [41]. The dielectric response of the two systems is also different at their magnetic transitions: While the paramagnetic-to-cycloidal phase transition in GaV$_4$S$_8$ at $T_c \approx 13$ K only leads to a very small anomaly in $\varepsilon'(T)$ at best, in GeV$_4$S$_8$ the dielectric constant is strongly suppressed when the system becomes antiferromagnetic at $T_N \approx 15$ K [23,29,39]. These differences may be ascribed to the different magnetic states in the two cases, but the detailed mechanisms leading to these anomalies are still far from being understood.

Providing more data on lacunar-spinel systems is prerequisite to get better insight into the nature of the ferroelectricity and of the dielectric behavior at their JT and magnetic transitions. In the present work we investigate single-crystalline samples of GaV$_4$Se$_8$. This material is expected to exhibit a similar orbitally-driven ferroelectric transition as the two other systems. There are only few reports on this compound in literature and, to our knowledge, its dielectric properties have only been studied in a very recent work presenting polarization measurements and the dielectric constant at a single frequency [42], which does not allow for a detailed analysis of the orbital and charge dynamics upon its phase transitions. In Ref. [43], using polycrystalline samples a magnetic transition at 10 K was reported. Our investigations by specific heat, magnetic susceptibility, dielectric spectroscopy, and polarization measurements reveal two subsequent phase transitions at about 42 and 18 K, which we identify with the JT and magnetic transitions, similar to the other lacunar spinels. These transitions were confirmed by two recent works on GaV$_4$Se$_8$, reporting, in addition, the presence of a skyrmion-lattice phase in external magnetic fields [42,44]. We detect strong dielectric and polarization anomalies at the JT transition which indicate JT-driven polar order in this compound, most likely of ferroelectric nature. Moreover, we find a small polarization enhancement when the material becomes multiferroic at the magnetic transition and enters a cycloidal spin state [42,44].

## II. EXPERIMENTAL DETAILS

Polycrystalline GaV$_4$Se$_8$ was prepared by solid-state synthesis from pure elements. The single crystals were grown by the chemical-transport reaction method using polycrystalline powder as starting material and iodine as transport agent. The growth was performed in a two-zone furnace between 850 and 800 °C. The x-ray diffraction study of crushed single crystals performed at room temperature revealed a cubic spinel structure with $F$-$43m$ symmetry and lattice constant 10.141(1) Å.

The heat capacity was investigated in a Quantum Design PPMS setup for temperatures between 2 and 60 K. Magnetic measurements were performed with a SQUID magnetometer (Quantum Design MPMS XL) in the temperature range from 2 to 60 K and in an external magnetic field of 1 T. For the dielectric and polarization measurements, contacts of silver paint were applied to opposite faces of platelet-shaped single crystals of about 0.25-0.9 mm$^2$ cross section and 0.45-0.71 mm thickness, ensuring an electric field applied along the <111> direction. The dielectric constant and conductivity were determined using the Keysight E4980AL Precision LCR Meter between 20 Hz and 1 MHz. Data between 4.4 K and room temperature were taken in a $^4$He-bath cryostat (Cryovac). To investigate the ferroelectric polarization, the pyrocurrent was measured as function of temperature between 2 and 58 K (cooling rate 5 K/min) using a Keithley 6517 electrometer. Sample cooling for these measurements was achieved by the PPMS.

## III. RESULTS AND DISCUSSION

Figure 1 shows the specific heat of GaV$_4$Se$_8$. Two peaks are observed at $T_{JT} \approx 42$ K and $T_m \approx 18$ K. They likely signify the JT and magnetic transitions, respectively, as expected based on the behavior documented for other lacunar spinels [23,25,29,32]. The spike-shaped peak detected at $T_{JT}$ indicates latent heat due to a first-order nature of the JT transition. For GaV$_4$S$_8$ and GeV$_4$S$_8$, such a first-order character was recently rigorously demonstrated by performing large-pulse heat-capacity measurements [29,39]. Because of the strong first-order character of the transition, we cannot provide a reasonable estimate of the associated entropy change as, e.g., contributions from latent heat could be easily underestimated [29].

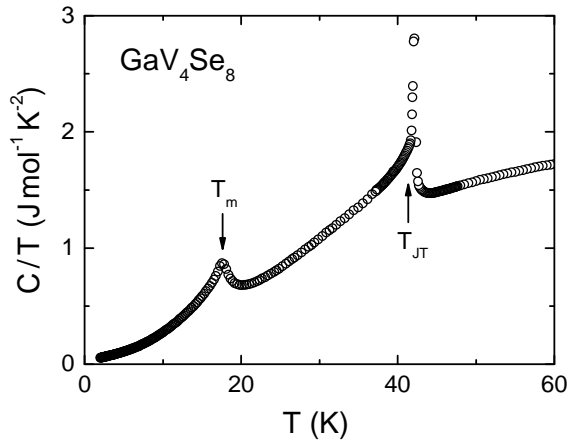

FIG. 1. Temperature-dependent specific heat of GaV$_4$Se$_8$ measured upon cooling (plotted as $C/T$ vs $T$ for clarity reasons). Two phase transitions at $T_m \approx 18$ K and $T_{JT} \approx 42$ K are revealed.

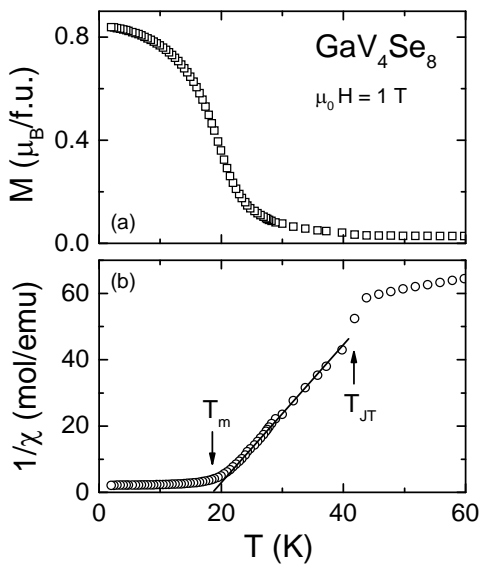

FIG. 2. Temperature-dependent magnetization (a) and inverse magnetic susceptibility (b) of GaV$_4$Se$_8$ measured under cooling. The line in (b) indicates Curie-Weiss behavior above the magnetic phase transition.

To verify the magnetic nature of the transition at $T_m$, in Fig. 2 the magnetization $M$ measured in an external field of 1 T and the inverse magnetic susceptibility $1/\chi$ are plotted. Starting at about 30 K, $M(T)$ [Fig. 2(a)] exhibits a strong increase with decreasing temperature signifying spontaneous magnetization, with maximum slope close to $T_m \approx 18$ K as determined by the specific-heat experiment (Fig. 1). At low temperatures it saturates, approaching a value of about 0.84 $\mu_B$. Moreover, the susceptibility, plotted as $1/\chi$ vs. $T$ in Fig. 2(b), reveals a Curie-Weiss behavior, i.e., $1/\chi \propto T - T_{CW}$, with $T_{CW} \approx 19$ K, again not far from $T_m$ deduced from specific heat. Thus, these results signify a transition of ferromagnetic nature close to $T_m$. Indeed, the saturation value of $M(T)$ seen in Fig. 2(a) is close to 1 $\mu_B$ as expected for a spin 1/2 system. This is consistent with the electronic structure of the V$_4$ clusters, with a single electron in their 3-fold degenerate highest occupied level [33]. These results were recently nicely confirmed by similar measurements reported in Ref. [42].

One should be aware, however, that in the closely related GaV$_4$S$_8$ a complex $H$-$T$ phase behavior was found for rather low fields (< 150 mT), with a cycloidal phase arising below about 13 K at zero magnetic field and ferromagnetic spin order showing up at elevated fields only [39]. For GaV$_4$Se$_8$ the situation indeed is similar and the transition at $T_m$, revealed for $H = 0$ in the specific heat data (Fig. 1), is of cycloidal nature [42,44], while it is clear from Fig. 2(a) that the ground state is turned to a fully polarized ferromagnet by as low fields as 1 T. However, one should note that the dominating exchange interaction still is ferromagnetic and possible deviations from ferromagnetism in zero field are due to secondary interactions only. Recently, extensive magnetic-field-dependent investigations of this system were performed to clarify the details of the phase diagram of GaV$_4$Se$_8$ [42,44] which, however, is out of the scope of the present work.

In addition to the anomaly at $T_m$, Fig. 2(b) reveals a step-like behavior of the magnetic susceptibility at $T_{JT}$. It signifies the structural transition at this temperature. Similar behavior was also observed in the sister compound GaV$_4$S$_8$ [38,39].

Figure 3(a) presents the temperature dependence of the dielectric constant at $T < 70$ K, as measured at various frequencies. At higher temperatures (not shown), $\varepsilon'$ exhibits a strong step-like and frequency-dependent increase, whose onset is observed at $T > 50$ K for the lower frequencies in Fig. 3(a). Analogous behavior was found for GaV$_4$S$_8$ (see Supplemental Material of Ref. [25]). It is due to a Maxwell-Wagner relaxation dominating the dielectric response at temperatures up to room temperature, which arises from contact effects and is a well-known non-intrinsic effect for semiconducting samples [45,46]. Similar to the findings in GaV$_4$S$_8$ [25], at low temperatures and sufficiently high frequencies it no longer contributes to the dielectric properties [46], which allows for the detection of the intrinsic dielectric properties.

For all frequencies, the $\varepsilon'(T)$ curves reveal well-pronounced and rather sharp peaks at the JT transition close to 42 K [Fig. 3(a)]. This finding strongly points to a JT-induced polar transition just as found for GeV$_4$S$_8$ and GaV$_4$S$_8$ [23,25]. The presence of a dielectric anomaly at this temperature was recently corroborated by $\varepsilon'(T)$ data measured at a single frequency of 10 kHz reported in [42]. While the main frame shows results collected upon cooling, the inset of Fig. 3(a) compares the cooling and heating curves for a single frequency of 1.58 kHz. A temperature hysteresis of about 1 K is revealed, in accord with the first-order character of the JT transition suggested above. At temperatures above and below $T_{JT}$, significant frequency dispersion of $\varepsilon'(T)$ is observed [Fig. 3(a)]. While at $T > T_{JT}$ an influence of the Maxwell-Wagner relaxation cannot be fully excluded, below the transition relaxational behavior as mostly found in order-disorder ferroelectrics [40] is likely responsible for this dispersion. For GaV$_4$S$_8$ such relaxation effects indeed were reported, although arising at higher frequencies than those found in the present work [47].

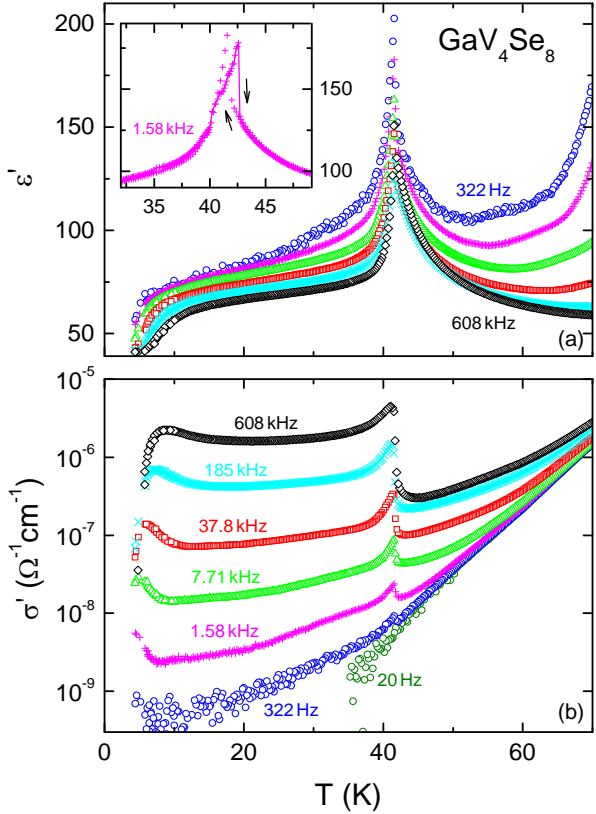

FIG. 3 (color online). Temperature dependence of dielectric constant (a) and conductivity (b) of GaV$_4$Se$_8$ measured at various frequencies under cooling. The inset shows both the cooling and heating curves of $\varepsilon'$ for 1.58 kHz (the lines connect the symbols).

Figure 3(b) shows the temperature dependence of the real part of the conductivity $\sigma'$ at different frequencies. It should be noted that, due to the relation $\sigma' \propto \varepsilon'' \nu$, the $\sigma'(T)$ curves in Fig. 3(b) also provide information on the temperature dependence of the dielectric loss $\varepsilon''$. The conductivity (and, thus, the loss) also exhibits a peak anomaly at the JT transition, similar to GaV$_4$S$_8$ [39]. The frequency dispersion of $\sigma'$ observed in Fig. 3(b) could partly be due to hopping conductivity, often leading to an increase of $\sigma'$ with frequency [48,49]. For lacunar spinels, in various works hopping conduction of electrons between the rather widely separated $M_4$ clusters was considered [33,50,51,52], which, however, is out of the scope of the present work. It should be noted that the mentioned anomaly in $\sigma'(T)$ at $T_{JT}$ diminishes with decreasing frequency and seems to be absent for the lowest measured frequency of 20 Hz. $\sigma'(T)$ at this frequency can be regarded as a good estimate of the dc conductivity.

The dielectric constant and conductivity documented in Fig. 3 do not reveal a direct signature of the magnetic transition close to 18 K. This resembles the behavior found in GaV$_4$S$_8$ when entering the cycloidal phase but is in contrast to that in GeV$_4$S$_8$ where $\varepsilon'(T)$ becomes strongly suppressed when the system becomes antiferromagnetic [23,25,29,39]. Instead of an anomaly fixed at $T_m$ for different frequencies as found at $T_{JT}$, $\varepsilon'(T)$ and $\sigma'(T)$ of GaV$_4$Se$_8$ exhibit *frequency-dependent* anomalies below $T_m$, namely a step and a peak, respectively. The onset of these features seems to arise just below $T_m$. However, to check this notion, measurements at higher frequencies would be necessary because these anomalies show a significant shift to high temperatures with increasing frequency. Interestingly, steps in $\varepsilon'(T)$ and peaks in $\sigma'(T)$ [or $\varepsilon''(T)$] behaving in this way are the typical signatures of a relaxational process that is continuously slowing down when temperature decreases. Being restricted to temperatures above 10 K, the measurements of $\varepsilon'(T)$ at 10 kHz reported in Ref. [42] did not probe this process.

To quantify this relaxational dynamics, Fig. 4 provides plots of the frequency-dependence of $\varepsilon''$ as measured at various temperatures in the region of the relaxational behavior revealed in Fig. 3(b). As expected for relaxation processes, peaks show up in $\varepsilon''(\nu)$. At the peak maxima, the frequency matches the relaxation rate of the relaxing entities and, via $\tau = 1/(2\pi\nu_p)$, the relaxation times $\tau$ can be determined from the peak frequencies $\nu_p$. The resulting $\tau(T)$ is plotted in the inset of Fig. 4 using an Arrhenius representation. The obtained linear behavior of log $\tau$ vs. $1/T$ indicates thermally activated relaxation dynamics with an energy barrier of 4.3 meV as determined from the linear fit shown by a solid line.

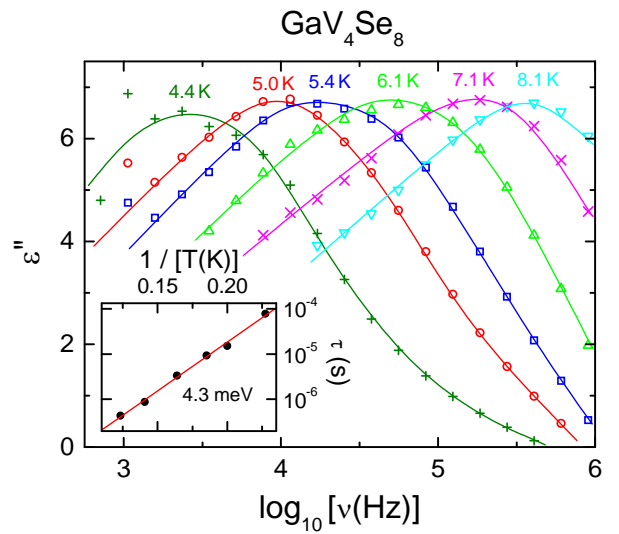

FIG. 4 (color online). Frequency dependence of the dielectric loss of GaV$_4$Se$_8$ for various temperatures showing the relaxational low-temperature behavior. The lines are guides to the eyes. The inset shows an Arrhenius plot of the temperature-dependent relaxation times as determined from the peak frequencies. The line is a fit with Arrhenius behavior, leading to an energy barrier of 4.3 meV.

Currently, we can only speculate about the microscopic origin of the relaxational process found at low temperatures in GaV$_4$Se$_8$. It may be related to the same dipolar degrees of freedom that also lead to the polar order below $T_{JT}$. For canonical order-disorder ferroelectrics, relaxational frequency dispersion is expected in the vicinity of the transition temperature [40]. In the present case of a JT-driven transition, electronic and ionic degrees of freedom are involved in the polar ordering [23]. When assuming only their partial coupling, the frequency dispersion found somewhat below $T_{JT}$ may arise from ionic displacements while the second relaxational dispersion region could be due to the faster electronic degrees of freedom that completely freeze only at much lower temperatures. Another possible origin of this low-temperature relaxation is the coupling between the spin and orbital (or lattice) degrees of freedom. Here we note that a similar low-temperature relaxation process is not present in GaV$_4$S$_8$ and GeV$_4$S$_8$ [25,29,39]. This peculiarity of GaV$_4$Se$_8$ should be addressed by future THz spectroscopy and/or magnetic-resonance spectroscopy measurements.

of self-aligned domains to guarantee a strong pyrocurrent signal even without poling. The very narrow shape of this peak also speaks against a TSC origin, which usually leads to much broader pyrocurrent peaks (see, e.g., [53]). The corresponding steplike increase in $P(T)$ [Fig. 5(b)] is very sharp, in agreement with the narrow peak in the specific heat (Fig. 1), again pointing to the first-order character of this transition. Below $T_{JT}$ the polarization saturates at about 1.0 μC/cm$^2$, which is of similar order as the values found for GaV$_4$S$_8$ [25] and GeV$_4$S$_8$ [23]. The emergence of polar order upon the JT transition, as already suggested by the appearance of a peak in the dielectric constant [Fig. 3(a)] is thus nicely confirmed by these polarization results.

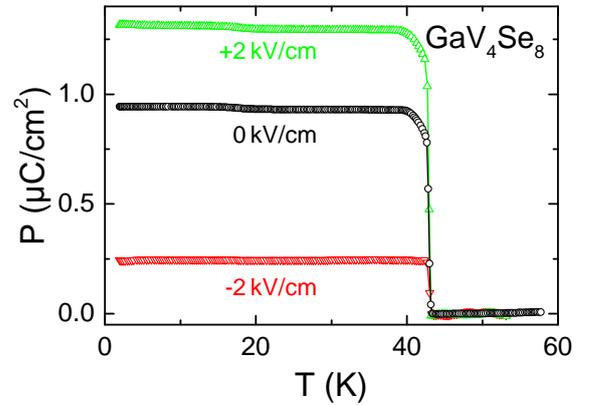

FIG. 6. Temperature dependence of the polarization of GaV$_4$Se$_8$ without poling and with opposite poling fields.

Overall, the present results on the temperature dependence of the dielectric constant [Fig. 3(a)] and polarization [Fig. 5(b)] reveal the typical signatures of ferroelectricity. However, the definition of ferroelectricity also includes the switchability of the polarization by an electrical field. This can be checked by measuring $P(E)$ hysteresis loops or by pyrocurrent measurements with positive and negative prepoling. Indeed, Singh et al. [23] demonstrated both for GeV$_4$S$_8$ using electrical fields up to 43 kV/cm. However, for GaV$_4$Se$_8$, due to electrical breakthroughs in our samples, we were not able to reach fields beyond about 2 kV/cm. Figure 6 shows polarization results as obtained from pyrocurrent measurements performed with positive and negative prepoling fields of 2 kV/cm. While a complete switching of $P$ into the opposite direction could not be achieved in this way, for the negative field at least a strong reduction of $P$ can be stated. This resembles the behavior found for GeV$_4$S$_8$ at limited fields [23] and may indicate at least partial switching of the polarization. We note here that, due to the lack of inversion symmetry in the high-temperature cubic state of lacunar spinels (space group: $F$-43$m$) and the formation of only one of the +/-P inversion domains upon the cubic-to-rhombohedral transition, as observed for GaV$_4$S$_8$ [26], the full reversibility of the polarization is not expected. In any

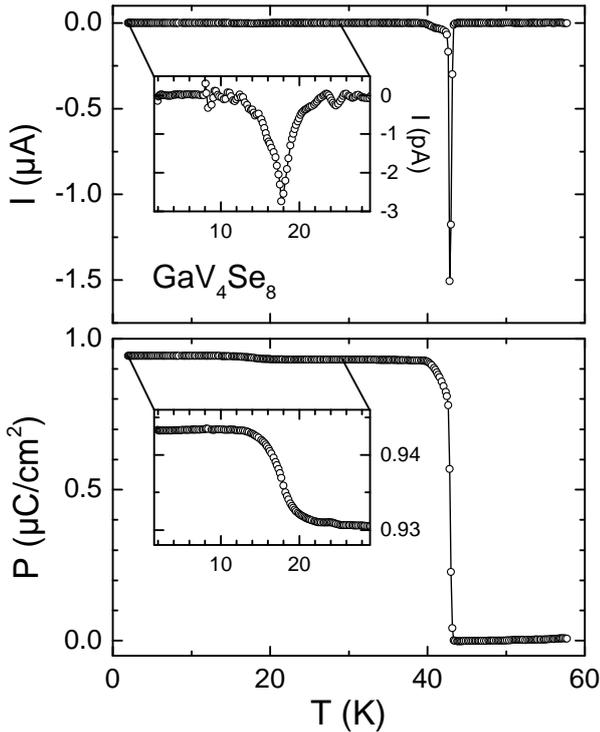

FIG. 5. Temperature dependence of the pyrocurrent (a) and the polarization (b) of GaV$_4$Se$_8$ without prepoling. The insets show zoomed views of the regions around the magnetic phase transition.

Figure 5(a) shows the pyrocurrent $I$ of GaV$_4$Se$_8$ as determined under heating. To exclude contributions from trapped charge carriers and thermally stimulated currents (TSC), these experiments were performed without previous poling of the sample. At the JT transition, a strong, spike-like negative peak occurs. Obviously, there are sufficient numbers case, we can unequivocally state the presence of a pyroelectric, polar state in GaV$_4$Se$_8$.

Interestingly, $P(T)$ in Fig. 5(b) also reveals a small but significant anomaly at the magnetic transition. This becomes obvious in the zoomed views provided by the insets of Fig. 5, where a small negative peak in $I(T)$ resulting in a somewhat smeared-out step-like increase of $P(T)$ with decreasing temperature is documented, arising close to $T_m \approx 18$ K. Very similar behavior of $P(T)$ at the magnetic transition with comparable magnitude was also found in GaV$_4$S$_8$ [25]. Obviously, just as in this system, in GaV$_4$Se$_8$ the polarization in the magnetic phase is slightly enhanced indicating some spin-driven contribution to $P$, in addition to the dominating JT-driven polarization. The coupling of magnetic order with electrical polarization in GaV$_4$Se$_8$ was recently confirmed by the magnetic-field dependent polarization measurements reported in Ref. [42]. In Ref. [25], the corresponding excess polarization in GaV$_4$S$_8$ was rationalized considering an exchange-striction mechanism.

## IV. SUMMARY AND CONCLUSIONS

In the present work, we have characterized the lacunar spinel GaV$_4$Se$_8$ by specific heat and magnetic susceptibility measurements and provided a detailed investigation of its dielectric and polarization properties. We find clear evidence for a polar transition occurring close to 42 K. The temperature dependence of the dielectric constant and polarization show the characteristics of a ferroelectric transition. We were able to demonstrate a strong variation of the polarization in external electrical fields up to 2 kV/cm, also consistent with a ferroelectric state, despite complete electric-field-induced polarization inversion could not be detected at these fields. We ascribe the polar transition to the same orbitally-driven mechanism as found in other lacunar spinels [23,25]. The dielectric anomaly at the transition markedly differs from the non-canonical behavior reported for GeV$_4$S$_8$ but it qualitatively resembles that found in GaV$_4$S$_8$, despite the ferroelectric transitions in all these materials are assumed to be orbitally driven. This difference most likely arises from the dissimilar electronic structure of the Ge compound compared to the two Ga systems: While in the latter two, a single electron occupies the uppermost triplet level of the V$_4$ clusters, in GeV$_4$S$_8$ there are two unpaired electrons with total spin $S = 1$ [27,28,33]. Thus the resulting JT transition differs from those in the two other compounds. Especially, the structure in the JT phase – orthorhombic for GeV$_4$S$_8$ and rhombohedral for the other two – and the distortion of the tetrahedra, contributing to the polar order, are not identical [28,31,33], rationalizing the different dielectric behavior at $T_{JT}$.

Moreover, we find clear indications for a relaxation process at low temperatures in GaV$_4$Se$_8$, below the magnetic phase transition. This relaxation may indicate partial quenching of orbital or lattice degrees of freedom, which are affected by the onset of the magnetic order via the spin-orbit or the spin-phonon coupling, respectively. Nevertheless, its origin remains to be clarified.

Finally, we detect a magnetic transition close to 18 K, which recently was shown to be of cycloidal nature [42,44]. Thus, GaV$_4$Se$_8$ exhibits simultaneous magnetic and polar order as also corroborated by the results of Ref. [42]. Thus, when assuming a ferroelectric nature of the latter, this compound can be regarded as type-I multiferroic with the rare realization of ferroelectricity combined with predominantly ferromagnetic spin alignment. Our polarization measurements provide evidence for spin-driven excess polar order when the material enters this magnetic phase, similar to previous findings in GaV$_4$S$_8$ [25]. Just as GaV$_4$S$_8$ [32], GaV$_4$Se$_8$ was recently reported to host Néel-type skyrmions [42,44], whirl-like topological spin objects that may be used for future data-storage applications. The skyrmions in this material also are dressed with electrical polarization [42], just as found for GaV$_4$S$_8$ [25], which makes this material even more interesting.

## ACKNOWLEDGMENTS

The authors thank S. Bordács for fruitful discussions and M. Csontos for technical assistance in the pyrocurrent measurements. This work was supported by the Deutsche Forschungsgemeinschaft through the Transregional Collaborative Research Center TRR 80 and by the Hungarian Research Fund OTKA K 108918.

―――――――――